\documentclass[twocolumn,trackchanges]{aastex631}
\usepackage{graphicx}
\usepackage{amsmath}
\usepackage{multirow}
\usepackage{array}
\usepackage[T1]{fontenc}
\usepackage[caption=false]{subfig}
\usepackage{ulem}
\usepackage{cancel}
\usepackage{threeparttable}
\usepackage{hyperref}  

\begin{document}

\title{Exploring the broadband spectral and timing characteristics of GRS 1915+105 with AstroSat and NICER observations}
\author{Ruchika Dhaka}
\affiliation{Department of Physics, IIT Kanpur, Kanpur, Uttar Pradesh
208016, India}

\author{Ranjeev Misra}
\affiliation{Inter-University Center for Astronomy and Astrophysics,
Ganeshkhind, Pune 411007, India}

\author{J.S. Yadav}
\affiliation{Department of Physics, IIT Kanpur, Kanpur, Uttar Pradesh
208016, India}
\affiliation{Tata Institute of Fundamental Research, Homi Bhabha Road, 400001, Mumbai, India}

\author{Pankaj Jain}
\affiliation{Department of Physics, IIT Kanpur, Kanpur, Uttar Pradesh
208016, India}

\begin{abstract}

In this study, we undertake a spectral-timing analysis of the black hole X-ray binary source GRS 1915+105 using simultaneous observations carried out by AstroSat (LAXPC and SXT) and NICER in 2017. The source showed two flux levels (high and low), whose energy spectra can be described by thermal comptonization of disk photons. The spectral parameters obtained by the joint fitting of SXT/LAXPC and NICER/LAXPC were consistent. The power density spectra from LAXPC and NICER revealed a broad, prominent feature at $\sim$2 Hz. The energy dependence of the fractional r.m.s and time-lag of this feature cannot be explained by only variations of coronal spectral parameters. Instead, a model where the coronal heating rate varies first and induces a change in the disk temperature and inner radius can explain the variation. Our results underline the importance of simultaneous observations by AstroSat and NICER and highlight the need for more sophisticated models to explain the spectral-temporal behavior of black hole systems.

\end{abstract}


\section{Introduction} \label{sec:intro}
GRS 1915+105, a Black Hole X-ray Binary (BHXB), came to the fore on 15 Aug 1992 and was detected by the WATCH All-Sky monitor on the Granat observatory. This binary system has a K-M spectral type donor star and a black hole with a mass equivalent to 12.4 solar masses and is located at a distance of 8.6 kpcs \citep{reid2014parallax}. This source has its relativistic jets inclined at an angle of $70^{\circ}$ from our line of sight \citep{mirabel1994superluminal}. This source is widely renowned for its diverse X-ray variabilities \citep{castro1992grs, belloni2000model,belloni1997unified, yadav1999different} and exhibits 14 distinct X-ray classes, differentiated based on its X-ray flux, Color-Color Diagram (CCD), and hardness ratio \citep[][]{belloni2000model,klein2002hard, hannikainen2005characterizing}. Since its first detection in 1992, GRS 1915+105 has primarily appeared in bright X-ray states like the High Soft State (HS) and High Hard Intermediate State (HIMS), among others. However, this source has transitioned into a diminishing phase in recent years, starting in 2018. 

X-ray binary systems exhibit rapid variability over short time scales. Fourier analysis is frequently used to understand rapid variations by computing power-density spectra (PDS) \citep[][]{van1989fourier}. Various patterns have been observed in PDS, spanning a spectrum of broadband noise patterns and more structured peaks known as Quasi-Periodic Oscillations (QPOs). For the source, GRS 1915+105, QPOs spanning frequencies from a few millihertz to up to approximately 70 Hz have been observed \citep[][]{ dhaka2023correlations, morgan1997rxte, belloni2013high, yadav1999different, paul1997quasi, sreehari2020astrosat}. QPOs are often characterized using a Lorentzian function, allowing us to assess their characteristics like centroid frequency ($f$), strength, and quality factor (defined as the ratio of the frequency of the QPO peak to its full width at half maximum). The centroid frequencies of these QPOs can be associated with physical processes occurring in these systems. Typically, two categories of QPOs exist. High-frequency QPOs (HFQPOs) have a centroid frequency of approximately 60 Hz or more (and in certain instances, up to a few hundred hertz) \citep[][]{belloni2012high,belloni2013discovery, belloni2009states} while Low-frequency QPOs (LFQPOs) exhibit a centroid frequency of around 30 Hz or less. Further, the LFQPOs have been classified into Types A, B, and C QPOs based on differences in power spectral properties and phase lag behavior \citep[][]{homan2001correlated, wijnands1999complex, remillard2002characterizing, casella2004study, ingram2019review}.

\begin{table*}[ht]
\centering
\scriptsize
\caption{Details of the source GRS 1915+105 observations by AstroSat and NICER. The table lists observation IDs alongside the instrument's name, exposure time, and observation date. \label{tab:table1}}
\begin{tabular}{lcccccc}
\hline
Observation No. & Telescope & Observation Date & Observation ID & Instrument & Exposure Time (ks) \\ 
\hline
1 & AstroSat & 31 Oct 2017 (MJD 58057) & G08\_028T01\_9000001656 & LAXPC \& SXT & 15.3 \& 5.5 \\  
2 & NICER & 31 Oct 2017 (MJD 58057) & 1103010132 & XTI & 4.9 \\ 
3 & NICER & 01 Nov 2017 (MJD 58058) & 1103010133 & XTI & 13.0 \\ 
\hline
\end{tabular}
\end{table*}

Numerous models have been proposed to explain the underlying dynamics of type-C QPOs. Some of these models center around geometric effects linked to the Lense-Thirring precession frequency, such as those discussed by \citet{stella1997lense}, \citet{schnittman2006precessing}, \citet{ingram2009low}. Other models suggest instabilities in the accretion flow \citep[][]{tagger1999accretion, cabanac2010variability}. \citet{misra2020identification}, \citet{liu2021testing}, \citet{dhaka2023correlations} have reported that the QPO might be linked to the dynamic frequency (denoted as $f_{Dyn} = c(s)/r$, with $c(s)$ representing sound speed) at the truncated disk radius. Various models of hydrodynamic origin are also found in works by \citet{chakrabarti2008evolution} and \citet{cabanac2010variability}. However, the definitive understanding of the physical origin of QPOs remains a subject of ongoing debate.  \\
The general behavior of BHXBs in terms of their spectral characteristics is typically described through a shift from a hard, less luminous state to a soft, more luminous state. This transformation is graphically represented as a q-shaped trajectory on the Hardness-Intensity Diagram (HID) \citep[][]{belloni2005evolution, homan2005evolution}. This trajectory encompasses intermediate regimes of hardness and luminosity. Notably, the frequency of type-C QPOs increases as the source makes transitions from harder to softer states. Depending on the theoretical framework, changes in hardness can be interpreted as variations in the outer radius of a hot inner flow \citet{esin1997advection}

Time lags between different energy photons are an important diagnostic to understand the nature of these systems. Reverberation lags, i.e., time-lags due to light travel time effects between the continuum photons and reflected features, can provide constraints on the geometry of the accretion disc \citep[][]{mastroserio2018multi, ingram2019public, lucchini2023investigating}. However, time lags between continuum photons require a different explanation. For kHz QPOs in neutron star systems, the time-lag is of the order of 50 microseconds, which is the expected timescale due to Compton scattering delays in a corona of size $\sim$10 km, i.e., the size of the neutron star \citep[][]{lee1998comptonization}. However, Compton scattering is expected to produce hard lags, while the observed ones were soft. \citet{lee2001compton} showed that soft lags are possible if a fraction of the comptonized photons impinge back into the soft photon source. This idea has been further developed in several works \citep[][]{kumar2016constraining, karpouzas2020comptonizing} for kHz QPOs in neutron star systems. This model has been extended by \citet{bellavita2022vkompth} to study low-frequency QPOs. However, since low-frequency oscillations show significantly longer time lags ($>$ 10 milliseconds), an explanation based on Compton scattering time scales requires large coronal sizes $>10^3$ km \citep[][]{bellavita2022vkompth, karpouzas2021variable, garcia2021two, zhang2022evolution}. 

\citet{mastichiadis2022study} describe a model where the C-type QPO is identified with the cooling time-scale of the corona. However, they invoke
an accretion rate, $\dot M \sim 10^{-3} L_{EDD}/c^2$, which with an efficiency of 10\% would lead to a luminosity of $\sim 10^{35}$ ergs/sec for a 10 solar mass black hole. In another interpretation, the C-type QPOs are identified as the inner hot corona precesing at the Lense-Thirring frequency \citep[][]{ingram2009low, ingram2016quasi}. The time lag will arise here since the observed spectral shape will change during different oscillation phases as the hot inner region angle with the disc varies. There is evidence that the shift in the iron line energy is a function of phase, which supports this model \citep[][]{ingram2009low, ingram2016quasi}.
However, for GRS 1915+105 analysis, \citet{nathan2022phase}, report that the phase-resolved analysis required a rather large disk thermalization time-scale to irradiation, and spectral analysis indicated a small inner disc radius incompatible with the Lense-Thirring interpretation. These interpretations are applicable for the particular case of C-type QPOs. 
Low-frequency broadband noise can be generated by stochastic variation in different radii of an accretion disc, which propagates inward, causing X-ray variation from the inner regions \citep[][]{lyubarskii1997flicker}. Such propagation can reproduce the observed frequency-dependent time-lags if the hard X-rays are preferentially produced in inner regions compared to the soft ones \citep[][]{kotov2001x}. While for a standard accretion disc, the higher energy photons are preferentially produced at smaller radii than low energy ones; this is unclear for thermal Comptonization photons. A high-energy thermal Comptonization photon has undergone several scatterings. Since the optical depth of the corona is typically of order unity, the photon would have scattered all over the corona. Hence, it cannot be associated with a particular corona region. In other words, it is unclear whether a corona can be divided into different zones, with the higher energy photons being produced preferentially in some zones compared to others. This situation may happen if the corona has large temperature gradients. Still, again, it is not clear whether this can happen; moreover, the time-averaged spectra usually assume a single-temperature corona.

It is important to find out under what conditions a single-zone compact thermal Corona can produce energy-dependent time-lags, which are larger than the scattering or light-crossing timescales. An approach is to consider the response of the energy spectrum to small variations in spectral parameters and then quantify which set of spectral parameters and time-lag between them can reproduce the observed energy-dependent fractional r.m.s and time-lag \citep[][]{maqbool2019stochastic, garg2020identifying}. The model assumes that at each phase of an oscillation, the energy spectrum can be represented by the same model used for the time-averaged one, with changes in the spectral parameters. The observed energy-dependent time lag is the consequence of any time lag between spectral parameters. Identifying the spectral parameters (and the time lag between them) responsible for the variation may lead to a better understanding of the temporal nature of the system. Ideally, the approach requires good broadband energy spectra and measurement of energy-dependent r.m.s. and time lag. 

In this study, we adopt the model proposed by Garg 2020 to understand the radiative components responsible for generating energy-dependent temporal variability in GRS 1915–105. We study the timing and spectral properties of GRS 1915+105 by using simultaneous data from AstroSat and NICER. We reviewed all the observations of GRS 1915+105 from AstroSat and NICER and selected the simultaneous observations. Our selection of simultaneous observations from AstroSat and NICER, with their broader energy coverage of 0.7–8.0 keV (SXT) and 3.0–80.0 keV (LAXPC) by AstroSat and 0.2–12 keV (XTI) by NICER, allows for better consideration of the disc emission. The high time resolution capabilities of NICER and LAXPC are especially advantageous for studying timing properties. This work utilizes the approach mentioned in \citet{garg2020identifying} and \citet{garg2022energy} to model and characterize the energy-dependent properties of the source.

Section \ref{sec:obs_data_red} outlines the observations used in this study, along with the data reduction methodologies employed for both NICER and AstroSat. Section \ref{sec:lc_hid} explains the lightcurve variation and HID. The study of spectral and temporal characteristics is reported in Section \ref{sec:spectral} and Section \ref{sec:timing}, respectively. Furthermore, in Section \ref{sec:timing}, we explain the model and the process of fitting timing characteristics, including fractional root mean square (fRMS) and time lag variability. Finally, in Section \ref{sec:conc}, we discuss the findings of this work.

\section{OBSERVATIONS AND DATA REDUCTION} \label{sec:obs_data_red}
\begin{figure*}
     \subfloat{
         \includegraphics[width=0.49\textwidth]{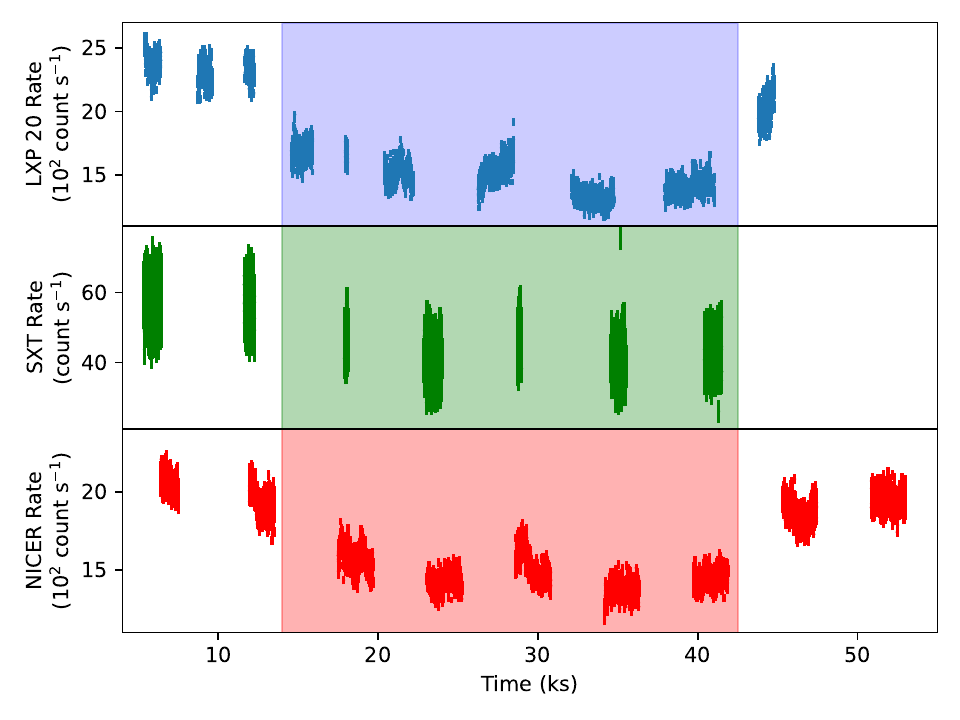}
         \label{fig:subfig1}
     }
     \hfill
     \subfloat{
         \includegraphics[width=0.49\textwidth]{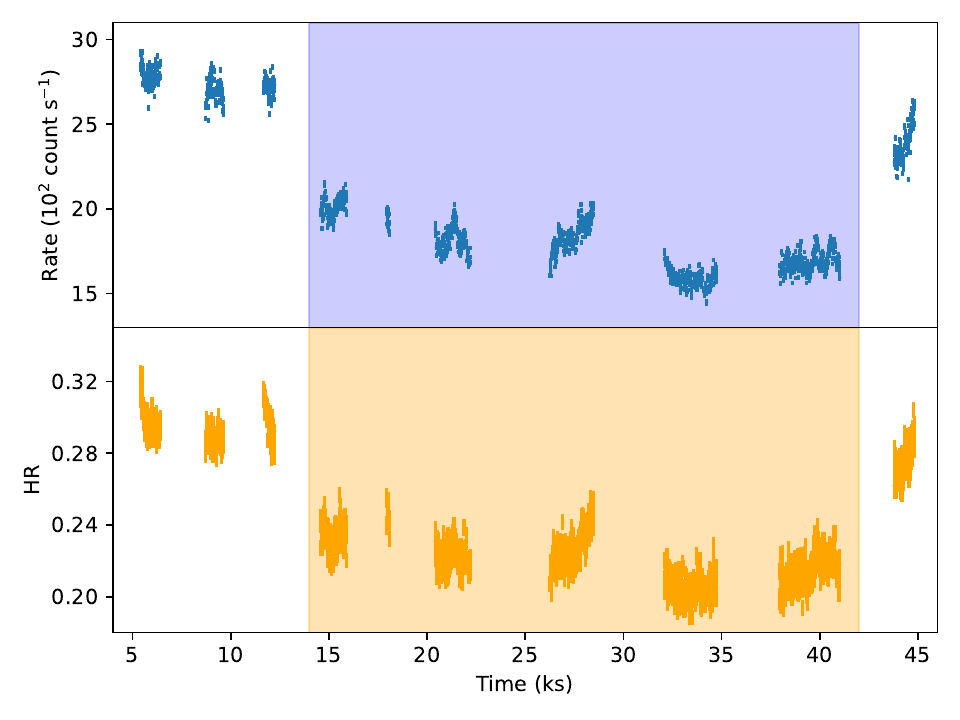}
         \label{fig:subfig2}
     }
     \caption{A 2.3775 binned background-subtracted light curve generated from LAXPC 20 in the energy range 4-50 keV (top left panel), SXT in energy range 1-7 keV (middle left panel) and NICER in the energy range 1.3-10.0 keV (bottom left panel) for the simultaneous observations made on 31 Oct and 1 Nov 2017. The right panel represents the LAXPC 20 lightcurve with the hardness ratio varying with time. The hardness ratio is defined as the ratio of the count rate in the 8-30 keV and 4-8 keV energy bands. The shaded region indicates the low flux regime, while the remaining area pertains to the high flux.}
\label{fig:fig1}
\end{figure*}

\subsection{Astrosat}
AstroSat \citep{agrawal2006broad}, is equipped with four co-aligned instruments that allow for the study of celestial objects in near and far-UV, soft X-rays (0.3-8 keV), and hard X-rays (3-100 keV). These instruments include the Ultra-Violet Imaging Telescope (UVIT) \citep[][]{tandon2017orbit}, the Soft X-ray Telescope (SXT) \citep[][]{singh2016orbit,singh2017soft}, the Large Area X-ray Proportional Counter (LAXPC) \citep[][]{yadav2016astrosat, antia2017calibration}, and the Cadmium-Zinc-Telluride Imager (CZTI) \citep[][]{bhalerao2017cadmium}. Notably, the LAXPC's large effective area and temporal resolution of 10$\mu$s provide a unique opportunity for detecting QPOs across a wide range of frequencies \citep[][]{yadav2016astrosat,chauhan2017astrosat}. For this work, we used publicly available AstroSat Guaranteed Time (GT) cycle data of GRS 1915+105. Table \ref{tab:table1} represents the AstroSat observations of GRS 1915+105, along with the effective exposure time and observation ID.

\subsubsection{Large Area X-ray Proportional Counter}

LAXPCs are comprised of three identical but independent PCUs (LAXPC 10, LAXPC20, and LAXPC30) with an effective area of 6000 cm$^2$ at 15 keV, a time resolution of 10$\mu$s in the energy range 3.0-80.0 keV and a dead-time of approximately 42 $\mu$s \citep[][]{yadav2016astrosat,yadav2016large,agrawal2017large}.

Level 2 event files were extracted from Level 1 event mode data utilizing the official LAXPC software version released on 15 Aug 2022\footnote{\href{http://astrosat-ssc.iucaa.in/laxpcData}{http://astrosat-ssc.iucaa.in/laxpcData}}.
For the temporal analysis of LAXPC, event files are barycenter corrected using AstroSat’s barycentric correction tool
as1bary\footnote{\href{http://astrosat-ssc.iucaa.in/?q=data\_and\_analysis}{http://astrosat-ssc.iucaa.in/?q=data\_and\_analysis}} to incorporate the effect of Earth and satellite motion relative to the barycenter of the solar system. The source spectra and light curves were obtained following the procedures outlined in \citet{agrawal2018spectral} and \citet{sreehari2019astrosat} for LAXPC data extraction and processing. Details regarding the response matrix (RMF) and generation of the background spectrum for proportional counters 10, 20, and 30  are provided in \citet{antia2017calibration}. Due to the gain instability issue caused by minor gas leakage, LAXPC30 is excluded \citep{antia2017calibration}. It should be noted that we extract the spectra from LAXPC20 data only because the gain remains stable throughout the observations. Both LAXPC 10 and LAXPC 20 PCUs have been used for timing analysis.

\subsubsection{Soft X-ray Telescope}

SXT is an imaging telescope that has a 2.3775 second time resolution. The Level 1 photon counting mode data of the SXT instrument was processed through the official.
SXT pipeline AS1SXTLevel2 - 1.4b\footnote{\href{https://www.tifr.res.in/~astrosat_sxt/sxtpipeline.html}{https://www.tifr.res.in/~astrosat\_sxt/sxtpipeline.html}} to produce Level 2 mode data. The Photon Counting mode (PC mode) data was chosen for the analysis. The HEASoft (version 6.29) tool XSELECT was used to generate the spectra, light curves, and images. The tools for SXT
data extraction, along with the background, spectral response, and
effective area files are provided by the SXT instrument team\footnote{\href{https://www.tifr.res.in/~astrosat_sxt/dataanalysis.html}{https://www.tifr.res.in/~astrosat\_sxt/dataanalysis.html}}. The sxtARFmodule\footnote{\href{https://www.tifr.res.in/~astrosat_sxt/sxtpipeline.html}{https://www.tifr.res.in/~astrosat\_sxt/sxtpipeline.html}} provided by the SXT instrument team was used to apply a correction for offset pointing. During the SXT observations, the count rate went over 40 counts per second in the High flux regime, As mentioned in the AstroSat Handbook\footnote{\href{https://www.iucaa.in/~astrosat/AstroSat_handbook.pdf}{https://www.iucaa.in/~astrosat/AstroSat\_handbook.pdf}}; this caused a pile-up effect at the center of the image due to the high flux of the source on the CCD. To tackle this, we set the inner radius of the circular annular region to 2 arcminutes. We generated light curves and spectra after excluding the central region of the image. No pile-up correction was necessary in the low flux regime, where the count rate was below 40 counts per second.

\subsection{Neutron Star Interior Composition Explorer}
Neutron Star Interior Composition Explorer (NICER) \citep[][]{gendreau2016neutron,gendreau2012neutron} is a payload
onboard the International Space Station (ISS). The X-ray Timing Instrument (XTI) of NICER comprises 56 X-ray optics with
silicon detectors operating in the 0.2–12 keV energy band \citep[][]{gendreau2016neutron}. Currently, 52 detectors are active. The NICER observed GRS 1915+105 from 31 Oct 2017 to 1 Nov 2017. 

We reduced the NICER data using the pipeline tool {\fontfamily{pcr}\selectfont nicerl2}\footnote{\href{https://heasarc.gsfc.nasa.gov/lheasoft/ftools/headas/nicerl2.html}{https://heasarc.gsfc.nasa.gov/lheasoft/ftools/headas/nicerl2.html}} in NICERDAS v11, available with HEASOFT v6.32, and applied the standard filters. We used the calibration database version xti20240206. Data from detectors 14 and 34, which were affected by increased electronic noise, were removed using the HEASOFT routine {\fontfamily{pcr}\selectfont fselect}. We use the FTOOL BARYCORR to apply the barycenter correction for each NICER observation.
Unprocessed event lists were obtained using {\fontfamily{pcr}\selectfont nicerl2} for both NICER observations. These lists were merged using {\fontfamily{pcr}\selectfont niextract-events}. The combined filter file was generated from the filter files of individual observation ID, using {\fontfamily{pcr}\selectfont ftmerge}, and the GTI for the combined observation was generated using {\fontfamily{pcr}\selectfont nimaketime}. Finally, {\fontfamily{pcr}\selectfont nicerclean} was used to produce a single cleaned event list for the two combined observation IDs. The source spectrum and the Scorpeon background file were extracted using the {\fontfamily{pcr}\selectfont nicerl3-spect} tool. Additionally, {\fontfamily{pcr}\selectfont nicerl3-spect} automatically applied a systematic error of about 1.5\% using {\fontfamily{pcr}\selectfont niphasyserr}. In this work, we used the NICER observations, simultaneous to AstroSat; their details are given in Table \ref{tab:table1}.

\section{Light curve and hardness-intensity diagram}
\label{sec:lc_hid} 
\begin{figure}
    \centering
    \includegraphics[width=1.\columnwidth]{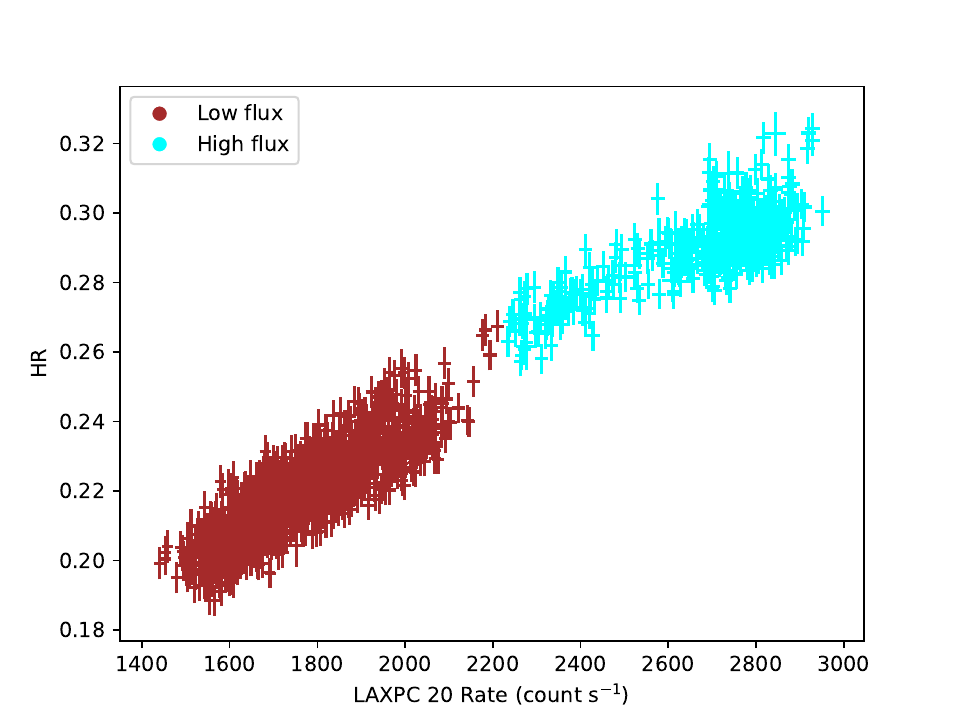}
   \caption{Hardness Intensity Diagram using LAXPC 20 data (observation ID G08\_028T01\_9000001656) with 10 sec time binning. The hardness ratio is consistent with the definition provided in Fig. \ref{fig:fig1}, and the X-ray intensity (X-axis) is extracted within the 4-30 keV energy range.}
    \label{fig:fig2}
\end{figure}

The top left panel of Figure \ref{fig:fig1} displays the background-subtracted lightcurve extracted from LAXPC 20 in the energy range of 4–50 keV. The time binning of the lightcurve is 2.3775 seconds to make it consistent with the time resolution of the SXT. The middle left panel shows the SXT light curve generated within the 1-5 keV energy interval. The bottom left panel shows the combined NICER light curve of observations 2 and 3. All the light curves are barycentric corrected. The right-hand panel of Figure \ref{fig:fig1} shows the LAXPC 20 light curve, with the variation of the hardness ratio over time. The hardness ratio is the count rate observed within the energy band of 8-30 keV to that within the 4-8 keV energy range. We partitioned the light curves into two distinct zones based on their X-ray flux and hardness intensity characteristics. The light curves exhibiting a count rate surpassing $\sim$2200 counts/sec (as depicted in the right panel of Figure \ref{fig:fig1}) were categorized as the high flux segment, whereas those with a flux falling below $\sim$2200 counts/sec were assigned to the low flux segment. The shaded part in Figure \ref{fig:fig1} represents the low flux regime. Throughout the manuscript, we shall consistently refer to these segments as the high and low flux regime light curves and corresponding spectra. Figure \ref{fig:fig2} illustrates the Hardness Intensity Diagram, generated using data from LAXPC 20, observation ID G08\_028T01\_9000001656. This diagram is made with a time binning of 10 seconds, and the X-ray intensity, presented on the X-axis, is extracted in an energy span of 4 to 30 keV. The two colors in Figure \ref{fig:fig2} correspond to low and high flux regimes.

\section{SPECTRAL ANALYSIS}
\label{sec:spectral}
\begin{table*}
\setlength{\tabcolsep}{4.4pt}
	\centering
	\caption{Broad-band X-ray spectral parameters for GRS 1915+105 using NICER/LXP20 and  SXT/LXP20 fitting for observations mentioned in table 1. The model used is {\fontfamily{pcr}\selectfont constant*tbabs*gabs*gabs(thcomp*diskbb+gaussian)}. (1) Exposure time of NICER/SXT in seconds; (2) Exposure time of LAXPC 20 in seconds; (3) Neutral hydrogen absorption column density in units of $10^{22} \textrm{cm}^{-2}$; (4)Thomson optical depth; (5) scattering fraction; (6) Temperature at inner disk radius in keV; (7) Normalization of diskbb; (8) iron line energy in keV; (9) line width in keV; (10) Normalization of Gaussian; (11) Strength of the first gabs component; (12) Strength of the second gabs component; (13,14,15) $\chi^2$ - statistics. The Gaussian absorption line energies were fixed at 6.97 keV and 8.25 keV, with line widths fixed at 0.05 keV in both {\fontfamily{pcr}\selectfont gabs} models. Columns two and three depict the spectral fitting of NICER/LXP20 for high and low flux, respectively, while columns four and five illustrate the spectral fitting of SXT/LXP20 for high and low flux, respectively.}
	\label{tab:table2}
	\begin{tabular}{c c @{\hspace{-1.0cm}} c c @{\hspace{-1.0cm}} c}  
		\hline
   & 
   
   \vspace{-0.3cm}\hspace{2.4cm}NICER+LAXPC & & 
   \hspace{2.4cm} SXT+LAXPC\\
   \multicolumn{1}{c}{Parameters} & \multicolumn{2}{c}{\hspace{1.35cm}\rule[0.01ex]{4.5cm}{0.5pt}\hspace{0.75cm}} & \multicolumn{2}{c}{\hspace{1.75cm}\rule[0.01ex]{4.5cm}{0.5pt}\hspace{0.75cm}} \\

	& High Flux & Low Flux & High Flux & Low Flux\\
		
		\hline
  Exposure Time NICER/SXT (sec)  &  7005 &  10870  &  1774 &  3718  \\
  Exposure Time LXP20 (sec) &  4089 &  11290 &  4089 &  11290  \\ [3pt]  
		$N_H$ ($10^{22} \textrm{cm}^{-2}$) & $4.27 \pm 0.02$ & $4.19 \pm 0.02$ &  $3.95 \pm 0.03$ & $3.85 \pm 0.04$ \\ [4pt]
		$\tau$  & $0.50^{+0.08}_{-0.12}$ & $0.42^{+0.04}_{-0.05}$ & $0.37^{+0.07}_{-0.14}$ & $0.33^{+0.01}_{-0.01}$ \\ [4pt]
            scattering frac.  & $0.4 \pm 0.2$ & $0.5 \pm 0.1$ & $0.79^{+0.11}_{-0.08}$ & $> 0.45$\\ [4pt]
            $T_{in} (keV)$ & $1.84 \pm 0.04$ & $1.52 \pm 0.03$ & $1.77 \pm 0.09$ & $1.34 \pm 0.02$ \\ [4pt]
            $Norm_{disk}$  & $137^{+10}_{-9}$ & $263^{+18}_{-14}$ & $148^{+31}_{-30}$ & $293^{+73}_{-60}$  \\ [4pt]
            
            $line E$  & $6.7^{+0.2}_{-0.4}$ & $6.6 \pm 0.1$ & $6.5^{+0.3}_{-0.4}$ & $6.4^{+0.2}_{-0.3}$ \\ [4pt]
            $line$ $width$($\sigma$) & $0.5^{+0.2}_{-0.2}$ & $0.7^{+0.2}_{-0.2}$  & $0.7^{+0.3}_{-0.2}$ & $0.5^{+0.3}_{-0.2}$  \\ [4pt]
             $Norm_{gauss}$  & $0.008 \pm 0.003$ & $0.017 \pm 0.005$ & $0.021^{+0.007}_{-0.009}$ & $0.015^{+0.006}_{-0.007}$  \\ [4pt]

     $strength_{gabs1}$  & $0.016 \pm 0.005$ & $0.028 \pm 0.005$ &  $< 0.04$& $<0.03$  \\ [4pt]

     $strength_{gabs2}$  & $0.031 \pm 0.005$ & $0.011 \pm 0.005$ & $0.11^{+0.14}_{-0.09}$  & $0.10^{+0.13}_{-0.09}$  \\ [4pt]
             \hline
            $\chi^2/$No. of bins (NICER/SXT) & 106/0/151 & 138.35/152  & 44.4/65
            &  46.7/67 
            \\ [1pt]
         $\chi^2/$No. of bins  (LAXPC 20) & 29.7/20 & 38.17/20 & 14.9/20 
         &  18.2/20  
            \\ [1pt]
          $\chi^2/$Dof (total) & 135.7/171 & 176.52/161  & 59.4/74 & 64.9/77 
          
          \\[1pt]
		\hline
	\end{tabular}
\end{table*}

We have used simultaneous NICER and LAXPC data fitting for the observations in Table \ref{tab:table1}. Using XSPEC 12.12.0, we performed a simultaneous spectral fitting of NICER and LAXPC 20 spectra.  
Then, to compare the combined analysis of NICER/LAXPC, we performed the combined fitting of SXT/LAXPC. We performed spectral fitting for the high and low flux regimes separately.
\begin{figure*}
     \centering
     \subfloat{
         \includegraphics[width=0.49\textwidth]{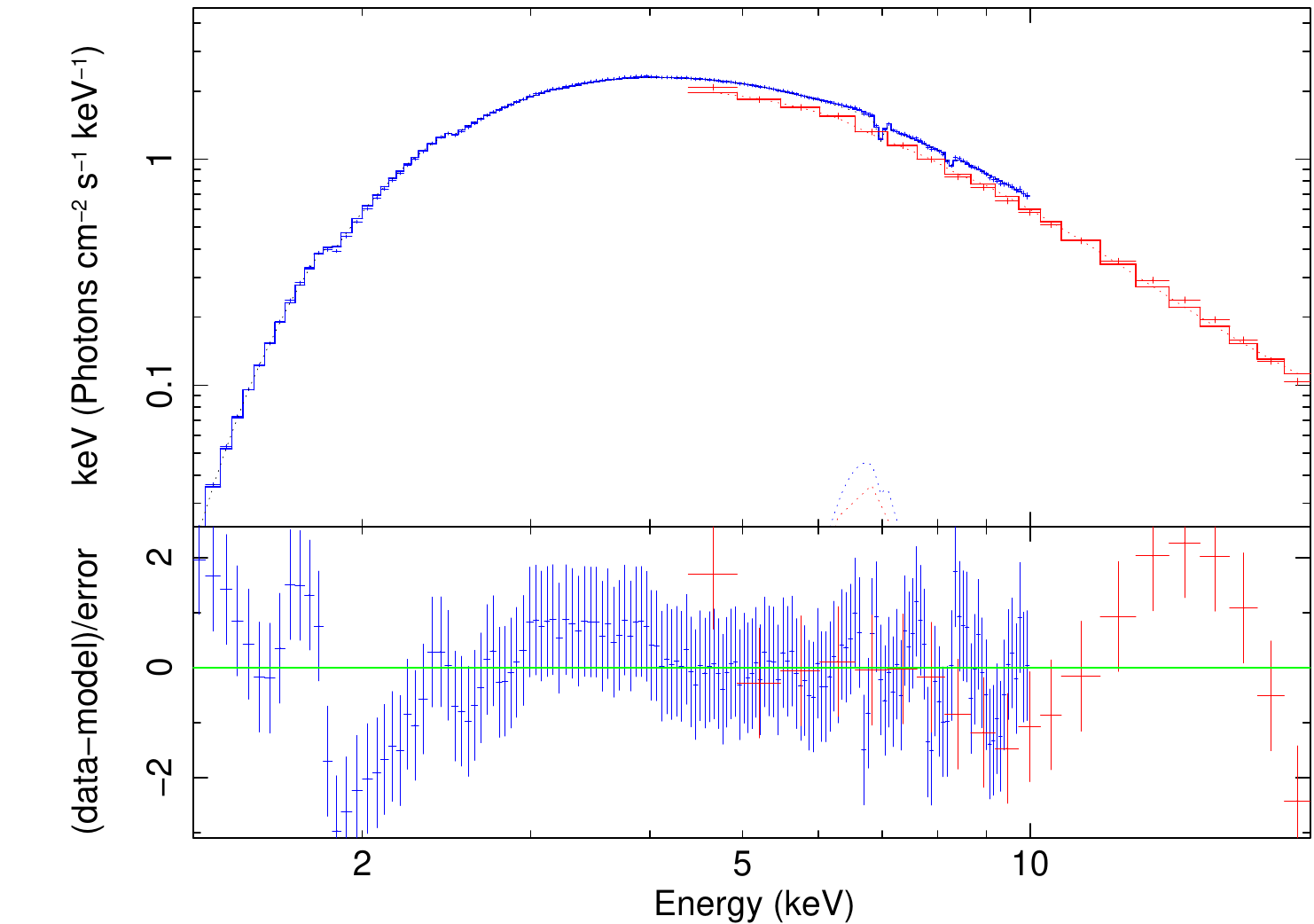}
         \label{fig:subfig3_1}
     }\subfloat{
         \includegraphics[width=0.49\textwidth]{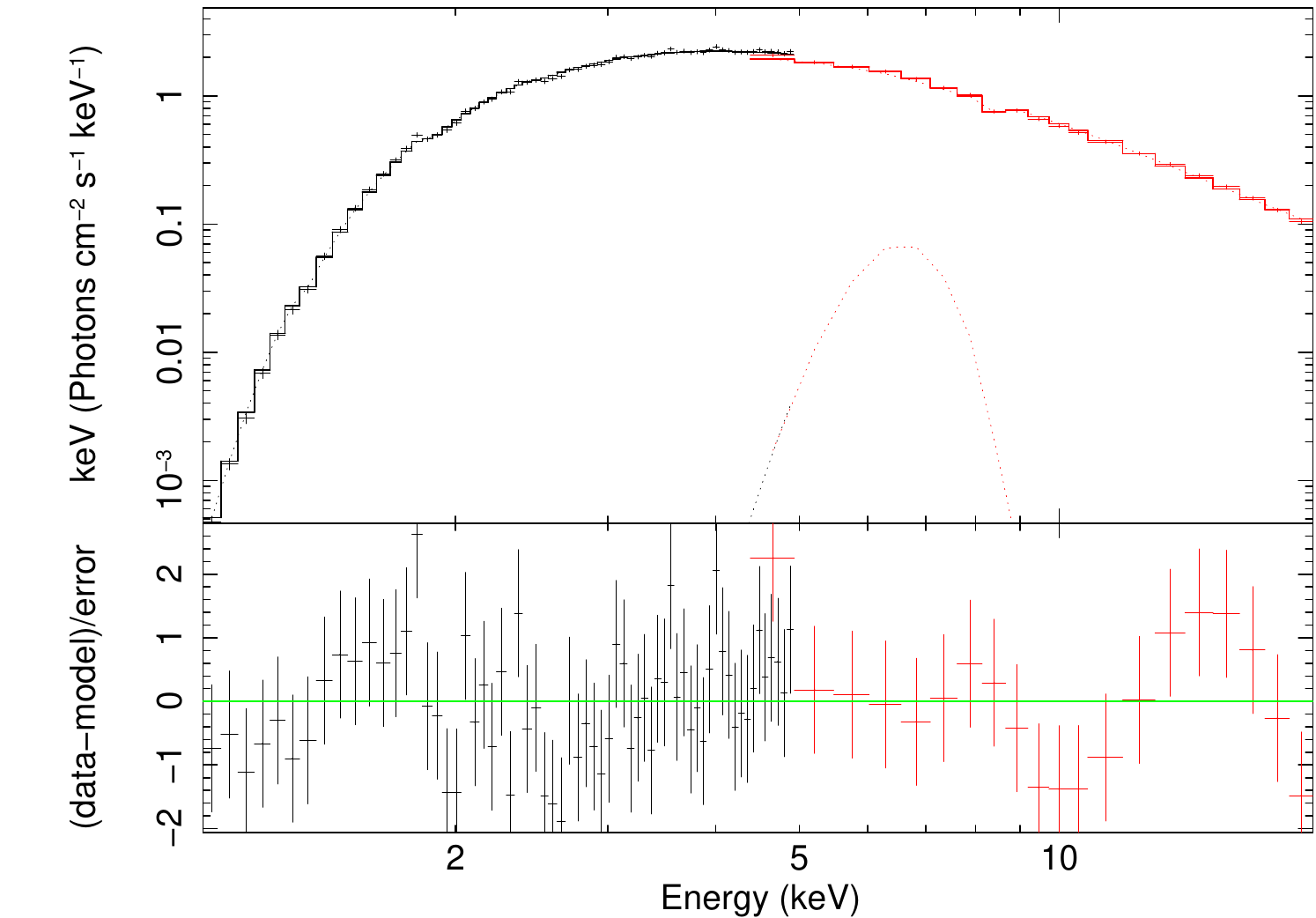}
         \label{fig:subfig3_2}
     }
     \par\medskip
     \subfloat{
         \includegraphics[width=0.49\textwidth]{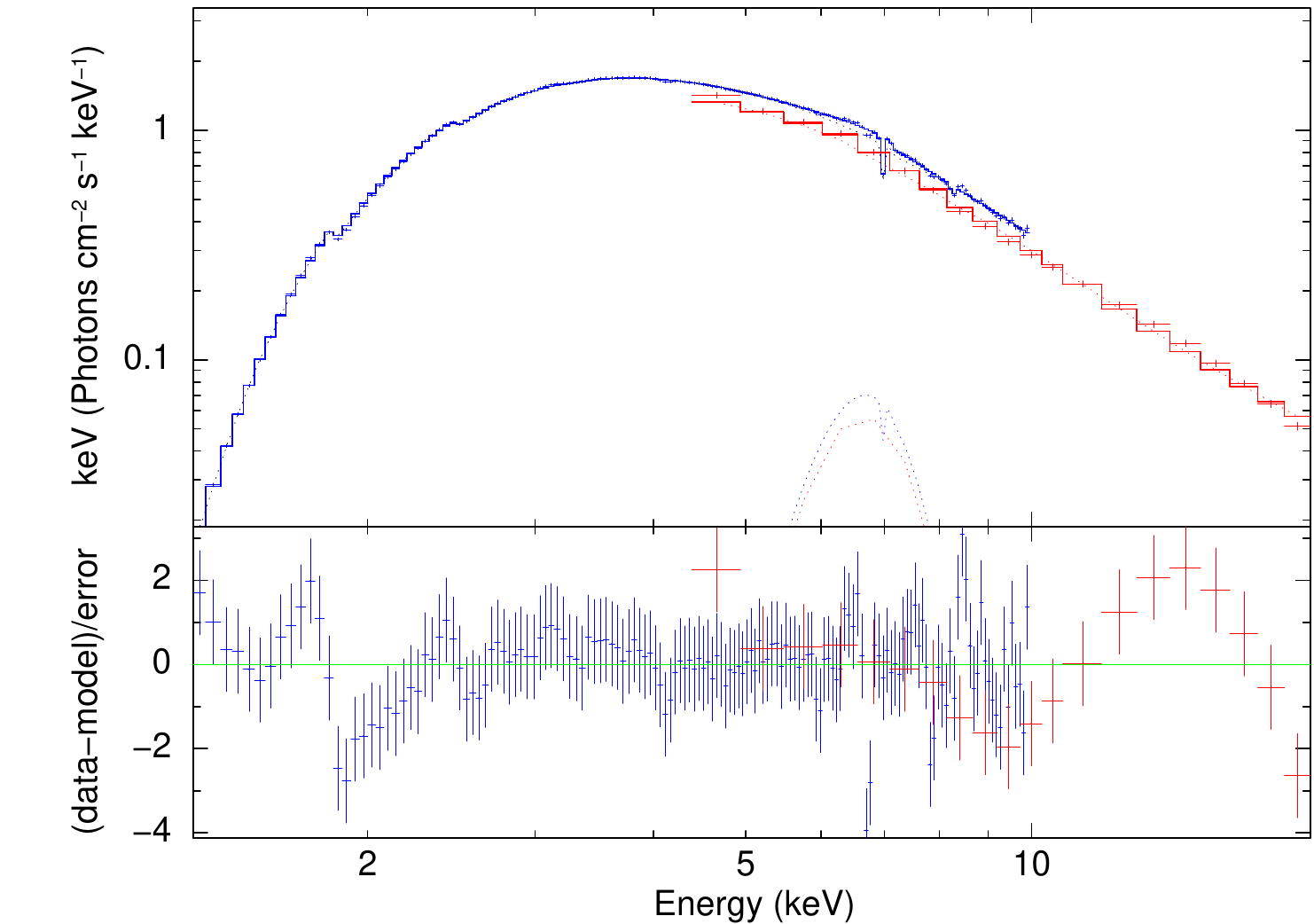}
         \label{fig:subfig3_3}
     }
     \subfloat{
         \includegraphics[width=0.49\textwidth]{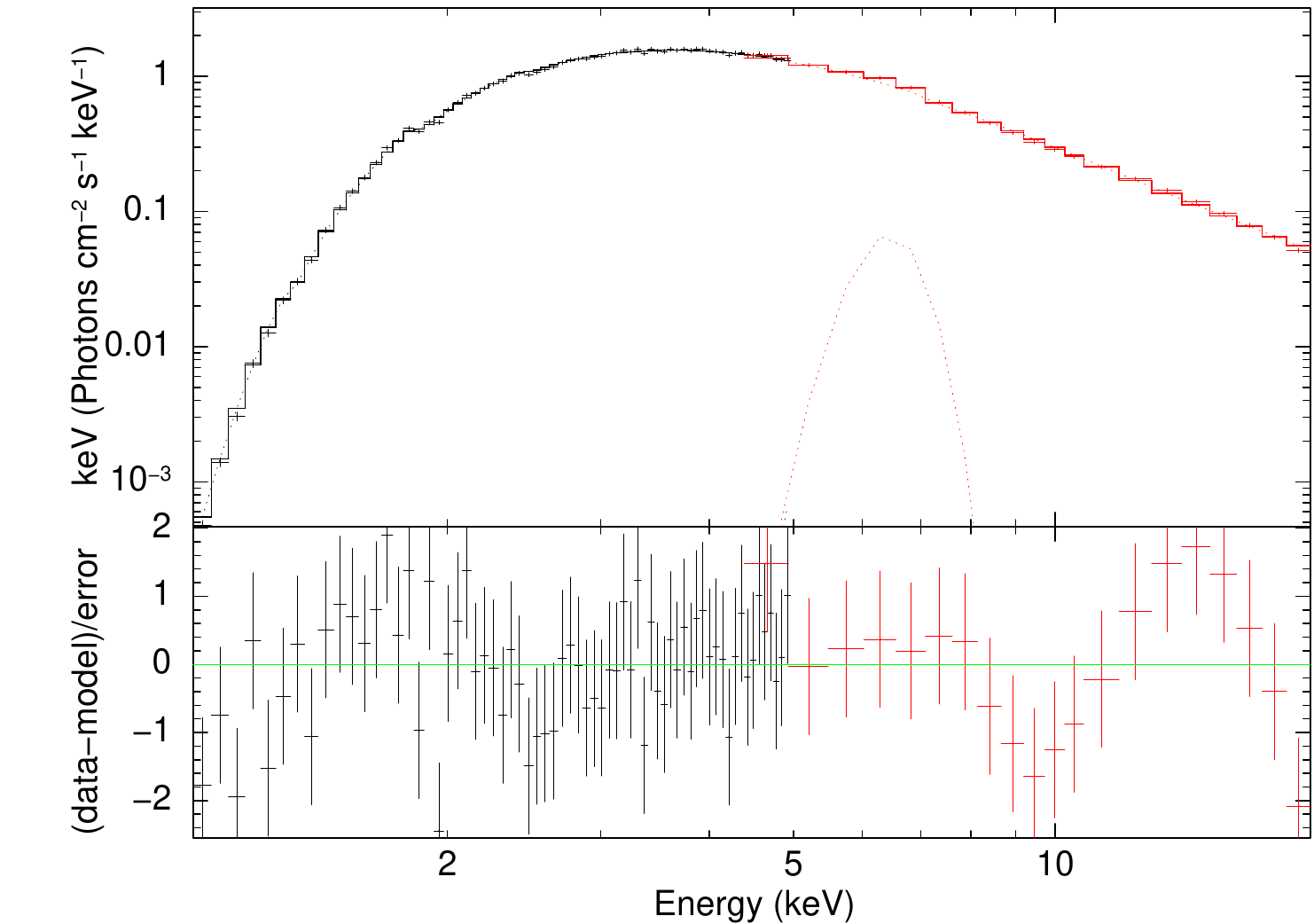}
         \label{fig:subfig3_4}
     }
    \caption{The left panels show the energy spectra of NICER/LAXPC of high flux in the top panel and low flux in the bottom panel, respectively, in the energy range of 1.5-20 keV. The SXT/LAXPC spectra for the high and low fluxes are shown in the right top and right bottom panels in the 1-20 keV energy range. The residuals are displayed beneath each spectrum. The data points for NICER, LAXPC 20, and SXT are represented in blue, red, and black, respectively.}
        \label{fig:fig3}
\end{figure*}

For the grouping of the spectra we used ftgrouppha\footnote{\href{https://heasarc.gsfc.nasa.gov/lheasoft/ftools/headas/ftgrouppha.html}{https://heasarc.gsfc.nasa.gov/lheasoft/ftools/headas/ftgrouppha.html}} tool of Ftools\footnote{\href{https://heasarc.gsfc.nasa.gov/ftools/}{https://heasarc.gsfc.nasa.gov/ftools/}} and used the optimal binning. To ensure consistency, we used a relative normalization constant for the simultaneous fitting of NICER/LAXPC and SXT/LAXPC data. Following the recommendation of the LAXPC and SXT team, a 3$\%$ systematic error was incorporated to account for uncertainties in the response of LAXPC and SXT instruments \citep[][]{antia2017calibration}. Due to inconsistencies in the effective area and response of the SXT, energies below 1 keV were not considered. Furthermore, a gain correction was applied to the SXT data, using the gain fit in XSPEC with a fixed slope of 1.

At low energies,  the NICER spectrum exhibits features likely caused by calibration uncertainties and the 'shelf' of the response from higher energy photons\footnote{\href{https://heasarc.gsfc.nasa.gov/docs/nicer/analysis\_threads/arf-rmf/}{https://heasarc.gsfc.nasa.gov/docs/nicer/analysis\_threads/arf-rmf/}}. This effect becomes more pronounced for highly absorbed sources like GRS 1915+105, where there are few detected source photons at low energies. Hence, \citet{nathan2022phase} and \citet{neilsen2020nicer} have limited their spectral analysis of NICER spectra to energies above 2.7 and 1.5 keV, respectively. Hence, We restrict the spectral fitting of NICER data to the 1.5 to 10.0 keV range. We performed a simultaneous spectral fitting of NICER and LAXPC spectra in the broad energy range of 1.5–20 keV.  \\
\begin{figure*}
\subfloat{
        \includegraphics[width=0.49\textwidth]{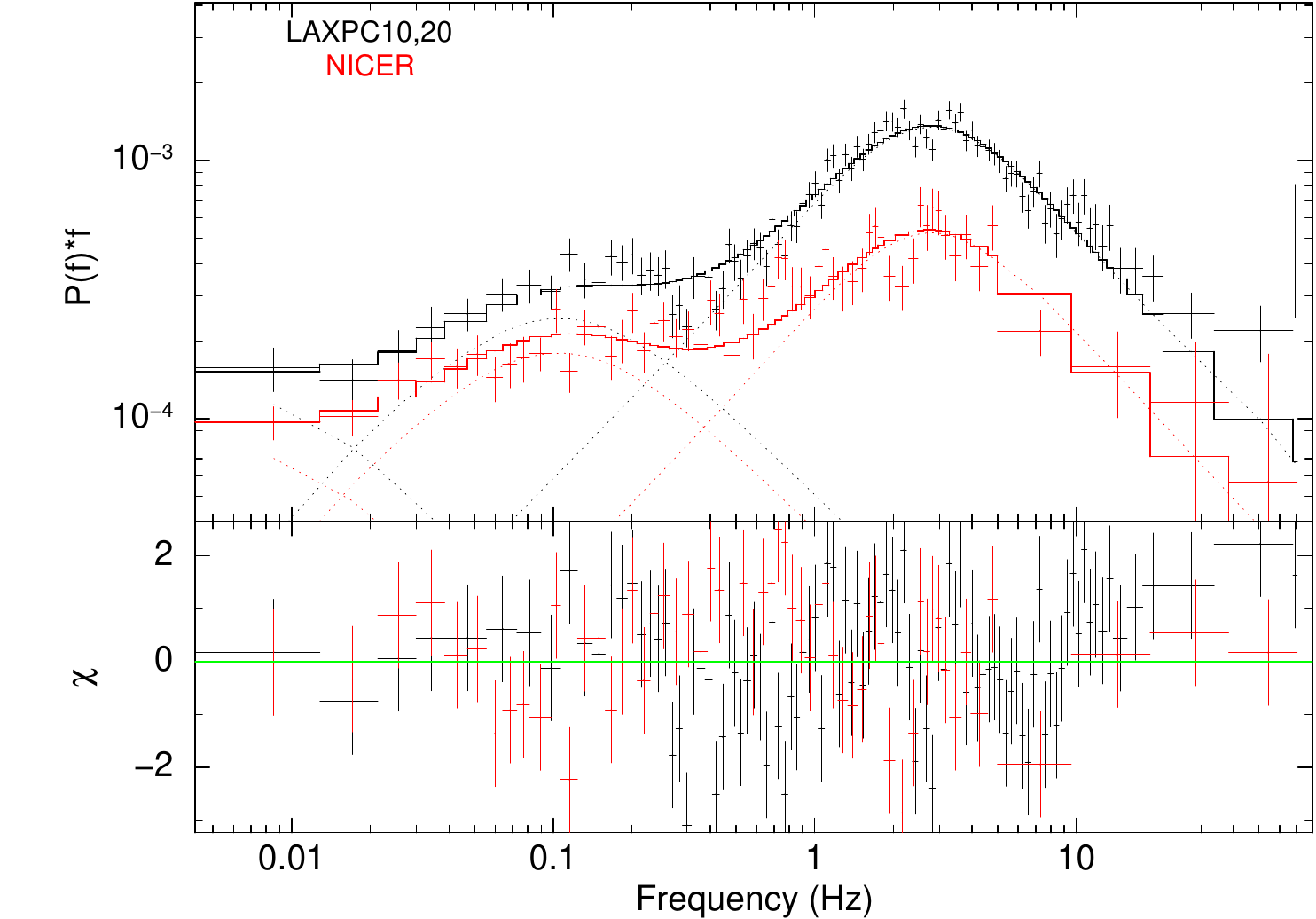}
        \label{fig:subfig4_1}
    }
    \hfill
    \subfloat{
        \includegraphics[width=0.49\textwidth]{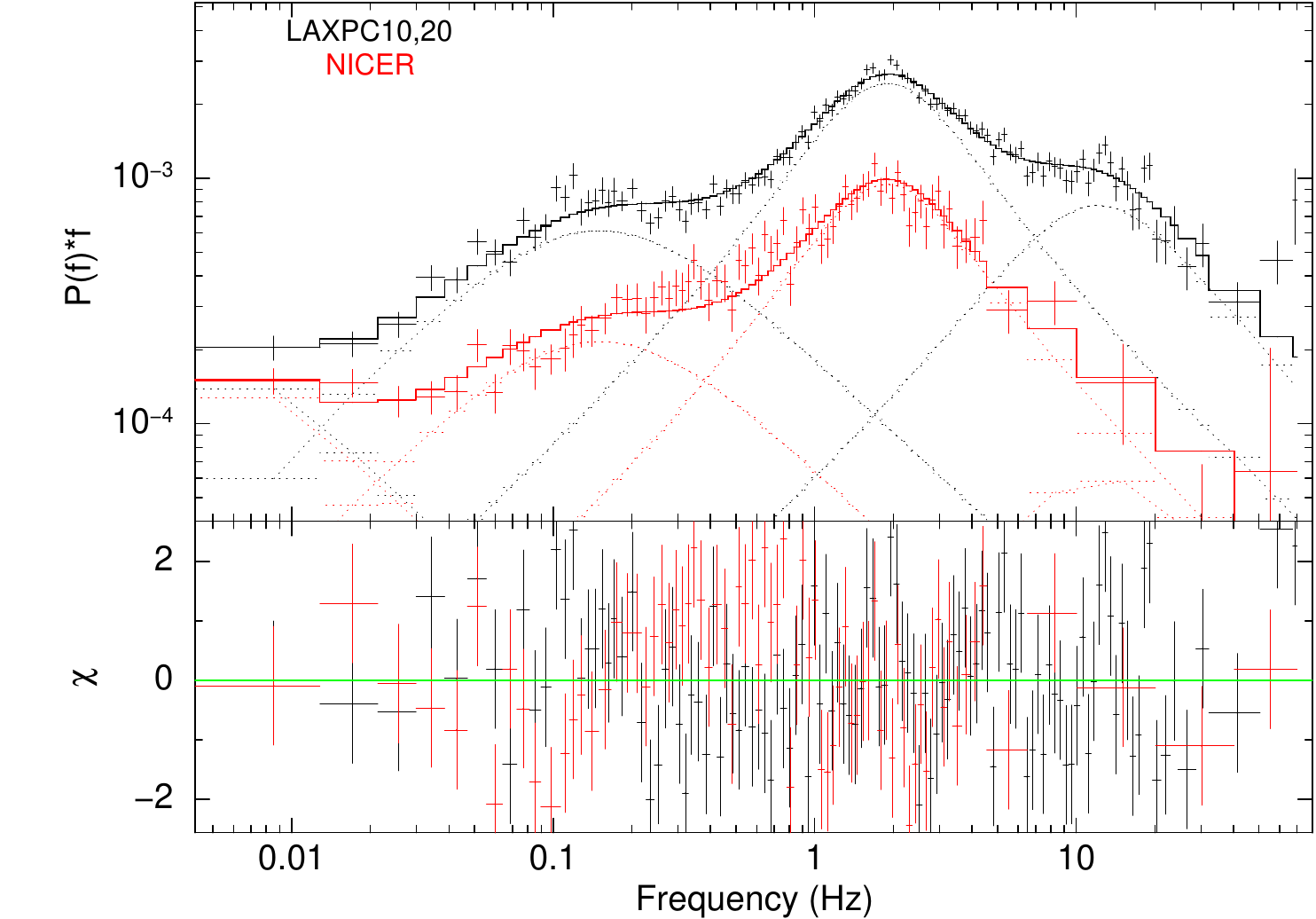}
        \label{fig:subfig4_2}
    }
    \caption{The left panel shows the power density spectra for high flux in the 0.005-80 Hz frequency range, using LAXPC 10,20 and NICER. The right panel shows the PDS of low flux in the frequency range of 0.005-80 Hz using LAXPC 10,20 and NICER. The red and black colors distinguish NICER and LAXPC 10, 20 PDS.}
\label{fig:fig4}
\end{figure*} \\
The spectra of BHXBs typically exhibit two main components: a multicolor blackbody disk emission and a high-energy thermal comptonization emission. To model the thermal disk emission, we used the {\fontfamily{pcr}\selectfont Diskbb} model \citep{mitsuda1984energy}, incorporating parameters such as inner disk temperature ($T_{in}$) and normalization ($N_{Disk}$). For the high-energy tail, we employed a Comptonization component, {\fontfamily{pcr}\selectfont thcomp} \citep{zdziarski2020spectral}, with parameters like photon index or Thomson optical depth, electron temperature, and scattering fraction. The coronal electron temperature was fixed at 100 keV, as it was not constrained, while all other parameters were allowed to vary. To account for the impact of interstellar medium (ISM) absorption, we used the {\fontfamily{pcr}\selectfont TBabs} model \citep{wilms2000absorption}. The GRS 1915+105 system exhibits complex absorption lines. Besides the broad iron line, there is evidence of a large number of narrow absorption lines. Lines corresponding to Iron K$\alpha$  at 6.4 keV, He-like Fe XXV at 6.7 keV, and H-like Fe XXVI at 6.97 keV have been observed in the spectra of this source. Additional lines are detectable at approximately 7.85 and 8.3 keV in certain intervals \citep[][]{nathan2022phase, neilsen2020nicer, wang20242022}. In the spectral fitting, we observed two absorption lines corresponding to Fe XXVI Ly$\alpha$ at 6.97 keV and Fe XXVI Ly$\beta$ at 8.25 keV and fitted them with Gaussian absorption models {\fontfamily{pcr}\selectfont gabs}. We freeze the width of the absorption lines at 0.05 keV and allow the strengths to vary. The top and bottom left panels in Figure \ref{fig:fig3} display the energy spectra of NICER/LAXPC in the high and low flux regimes, respectively. Residuals for each spectrum are presented below the corresponding spectrum.

To compare the simultaneous analysis of NICER/LAXPC, we performed the simultaneous fitting of SXT/LAXPC spectra across the energy range of 1-20 keV using the same model. Details of the spectral results of NICER/LAXPC and SXT/LAXPC are provided in Table \ref{tab:table3}, covering the parameters like hydrogen absorption, electron temperature, disc temperature, and scattering fraction. We have restricted the spectral analysis of LAXPC data to less than 20 keV since beyond that, there are slight broad systematic residuals at the 2-sigma level, which may be due to some complex spectral features or a slight underestimation of the background. \\

Having obtained a reasonable fit, we replaced the {\fontfamily{pcr}\selectfont diskbb} models with more physically motivated, namely the relativistic model {\fontfamily{pcr}\selectfont kerrd} \citep[][]{ebisawa2003accretion}. For the {\fontfamily{pcr}\selectfont kerrd} model, the black hole mass, disc inclination angle, and distance to the source were kept constant at values of 12.4$M_{\odot}$, 60$^{\circ}$ and 8.6 kpc, respectively \citep{reid2014parallax}. The spectral hardening factor was set at 1.7 \citep{shimura1995spectral}. We note that the mass accretion rate is \(1.37 \pm 0.04 \times 10^{18} \text{gm/s}\) for high flux and \(1.12 \pm 0.05 \times 10^{18} \ \text{gm/s}\) for low flux in the NICER/LXP20 spectra. In the SXT/LXP20 spectra, the mass accretion rate is \(1.35 \pm 0.07 \times 10^{18} \ \text{gm/s}\) for high flux and \(1.19 \pm 0.21 \times 10^{18} \ \text{gm/s}\) for low flux. The inner radius is \(<2.1\ R_{g}\) for high flux and \(2.2 \pm 0.2 \ R_{g}\) for low flux in the NICER/LXP20 spectra, while in the SXT/LXP20 spectra, it is \(<2.5 \ R_{g}\) for high flux and \(<2.7 \ R_{g}\) for low flux.

\section{Broad-band Timing ANALYSIS}
\label{sec:timing}
\subsection{Power Density Spectra (PDS)}
To understand the variability in the light curve, we study the frequency domain through the computation of PDS. To compute the PDS, we combine data from LAXPC detector units, LAXPC 10 and LAXPC 20, along with NICER data. Separate PDS extractions are performed for high and low flux regimes to discern differences in timing characteristics. To analyze the variability within the LAXPC data, we employ the \textit{laxpc\_find\_freqlag} subroutine for PDS computation in the energy range of 4-50 keV. The subroutine first generated a lightcurve in 4-50 keV energy band with a time resolution of $6.25 \times 10^{-2}$ sec. Subsequently, the lightcurves are divided into multiple segments with a $\sim$200-sec length. The PDS is calculated for each segment with a Nyquist frequency of $\sim$80 Hz. The resulting PDS segments are then averaged to produce the final PDS. Similarly, for NICER data, the \textit{nicer\_find\_freqlag} subroutine is utilized for PDS computation in the energy range of 0.8-10 keV. The dead-time corrected Poisson noise has been subtracted from the PDS, and the PDS normalization has been adjusted, taking the background rate into consideration \citep[][]{yadav2016astrosat}{}. We implemented a multi-Lorentzian model to fit the PDS \citep[][]{belloni2002unified, nowak2000there} and found that four components were required to fit the PDS for low and three for high flux regimes. We performed a simultaneous fitting of the LAXPC and NICER PDS. During this fitting, we tied each Lorentzian component's peak frequency and width in the NICER PDS to those of the LAXPC PDS. Additionally, we kept the normalization of all Lorentzian components free. Each Lorentzian is described with three parameters: centroid frequency ($f$), FWHM ($\sigma$), and normalization ($N_L$) of the respective components. Table \ref{tab:table3} shows the list of best-fitting parameters with errors estimated in 90 percent confidence region and RMS (percent) contribution for each component calculated by taking the square root of Lorentzian Norm. In the left panel, Figure \ref{fig:fig4}, we present the PDS for the high flux regime of both LAXPC and NICER within the frequency range of 0.005-80 Hz. Similarly, the right panel shows the PDS for the low flux regime of LAXPC and NICER within the same frequency range. The color scheme distinguishes between NICER and LAXPC data, with red and black representing NICER and LAXPC, respectively. No QPO is observed in either the high or low flux regimes of the PDS depicted in Figure \ref{fig:fig4}. Instead, broad features emerge at frequencies $\sim$0.1 Hz, $\sim$2 Hz, and $\sim$12 Hz. The best-fitting parameters of the second Lorentzian component (L2) in Table \ref{tab:table3} shows that the power is prominent at $\sim$2 Hz feature with quality factor; Q $\sim$0.26 for high flux regime and Q$\sim$0.49 for low flux regime. Meanwhile, the rest of the Lorentzian components describe the broader noise components in the PDS compared to L2. Hence, the power prominence at $\sim$2 Hz stands out notably compared to the other two broadband features. To check whether the broad feature at $\sim$2 Hz may be due to a time-varying narrow QPO, we computed and examined the dynamic power spectra. However, we found no evidence of a narrow, time-varying QPO. Consequently, we studied the root mean square (RMS) and time lag spectra at the $\sim$2 Hz frequency in detail. From now on, we will refer to this notable broadband feature as the one at $\sim$2 Hz. \\
\subsection{Energy dependent Fractional RMS and Time-Lags}
We studied the fractional root mean square (fRMS) and time-lag variation with energy to understand the source's energy-dependent characteristics. This analysis is focused on a broad feature at $\sim$2 Hz in the PDS. For the computation of time-lags, the complete methodology is outlined in \citet{nowak1999low}. 
\begin{table}
	\centering
	\caption{The best-fitting parameter values obtained by fitting the PDS within the 3–30 keV range for LAXPC and 0.3–12 keV for NICER are listed below. Three Lorentzian components were used to fit the PDS for the high flux regime, while four Lorentzian components were required for the low flux regime. The reported errors are presented with a confidence level of 90 $\%$.}
	\label{tab:table3}
	\begin{tabular}{lccr} 
		\hline
		Component & Parameters & High Flux & Low Flux \\
		\hline
L1 &  $f$ (Hz)  &  0.0($f$) &  0.0($f$)  \\    [4pt]
 &  $\sigma_{1}$  &  $0.008^{+0.013}_{-0.006}$ &  $0.007 \pm 0.006$   \\    [4pt]
LAXPC &  $N_{L1} (10^{-3})$ &  $0.4^{+0.2}_{-0.1}$ &  $0.6^{+0.2}_{-0.2}$   \\    [4pt]
 &  $fRMS_{L1}$ (\%) &  $2.0^{+0.5}_{-0.4}$ &  $2.5 \pm 0.4$   \\    [4pt]
NICER  &  $N_{L1} (10^{-3})$ &  $0.3^{+10.1}_{-0.1}$  &  $0.6^{+12.3}_{-0.2}$  \\    [4pt]
 &  $fRMS_{L1}$ (\%)  &  $1.7^{+31.0}_{-0.4}$ &  $2.4 \pm 0.4$ \\    [4pt]
\hline 
\vspace{-6pt} \\ [-6pt]
L2 &  $f$ (Hz) &  $1.2 \pm 0.2$ &  $1.33 \pm 0.06$   \\    [4pt]
 &  $\sigma_{2}$   & $5.1 \pm 0.2$  & $2.7 \pm 0.2$ \\    [4pt]
LAXPC &  $N_{L2} (10^{-3})$ & $3.5 \pm 0.1$ & $4.8 \pm 0.3$   \\    [4pt]
 &  $fRMS_{L2}$ (\%) & $5.9 \pm 0.1$  &  $6.9 \pm 0.3$ \\    [4pt]
NICER  &  $N_{L2} (10^{-3})$ &  $1.4 \pm 0.1$ &  $1.9 \pm 0.1$   \\    [4pt]
 &  $fRMS_{L2}$ (\%) &  $3.7 \pm 0.1$ &  $4.3 \pm 0.2$  \\    [4pt]
\hline 
\vspace{-6pt} \\ [-6pt]
L3 &  $f$ (Hz) & 0.0($f$)  &  0.0($f$) \\    [4pt]
 &  $\sigma_{3}$  &  $0.30^{+0.04}_{-0.03}$  &  $0.30 \pm 0.03$  \\    [4pt]
LAXPC &  $N_{L3} (10^{-3})$ &  $0.8 \pm 0.1$ &  $1.9 \pm 0.1$  \\    [4pt]
&  $fRMS_{L3}$ (\%)  &  $2.8 \pm 0.2$ &  $4.0 \pm 0.1$ \\  [4pt]
NICER  &  $N_{L3} (10^{-3})$ &  $0.6 \pm 0.1$  &  $0.7 \pm 0.1$  \\    [4pt]
 &  $fRMS_{L3}$ (\%) & $2.4 \pm 0.1$  &  $2.6 \pm 0.1$  \\    [4pt]
\hline   
\vspace{-6pt} \\ [-6pt]
L4 &  $f$ (Hz) &  ...  &  $6.6^{+2.5}_{-3.8}$ \\    [4pt]
 &  $\sigma_{3}$  &  ... &  $20.4 \pm 2.9$  \\    [4pt]
LAXPC &  $N_{L3} (10^{-3})$ &  ...  &  $1.2^{+0.3}_{-0.4}$  \\    [4pt]
&  $fRMS_{L3}$ (\%)  & ... &  $4.3 \pm 0.4$ \\  [4pt]
NICER  &  $N_{L3} (10^{-3})$ & ...  &  $0.2 \pm 0.2$  \\    [4pt]
 &  $fRMS_{L3}$ (\%) & ...  &  $1.2 \pm 0.8$  \\    [4pt]
\hline   
\vspace{-6pt} \\ [-6pt]
$\chi^2/$d.o.f. &  & 210.55/148 &  248.98/179  \\    [4pt]
\vspace{-6pt} \\ [-6pt]
\hline
\end{tabular}
\tablecomments{$f$ denotes a fixed parameter during the fitting.}
\end{table}  
\begin{figure*}
    \subfloat{
        \includegraphics[width=0.49\textwidth]{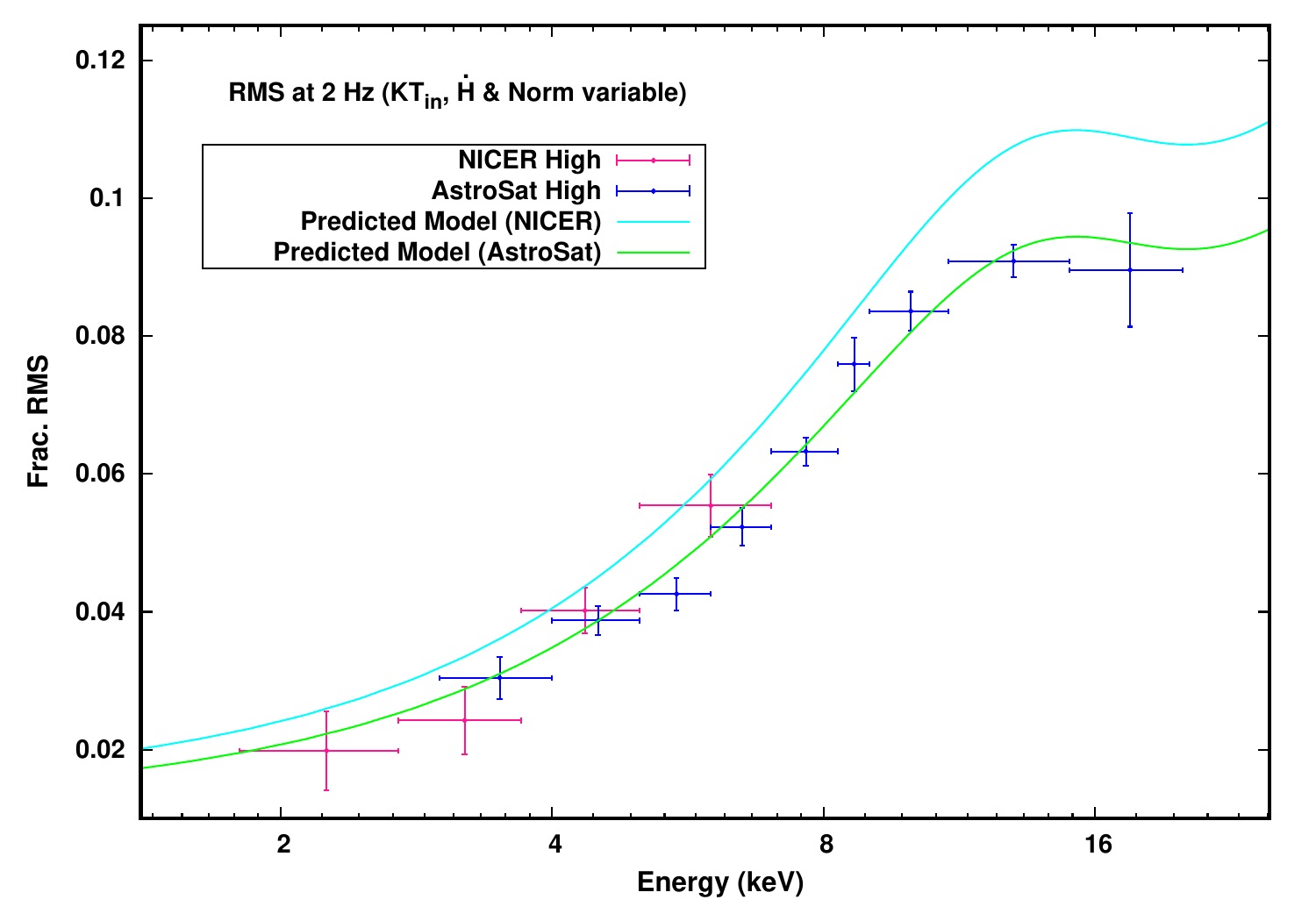}
        \label{fig:subfig5_1}
    }
    \hfill
    \subfloat{
        \includegraphics[width=0.49\textwidth]{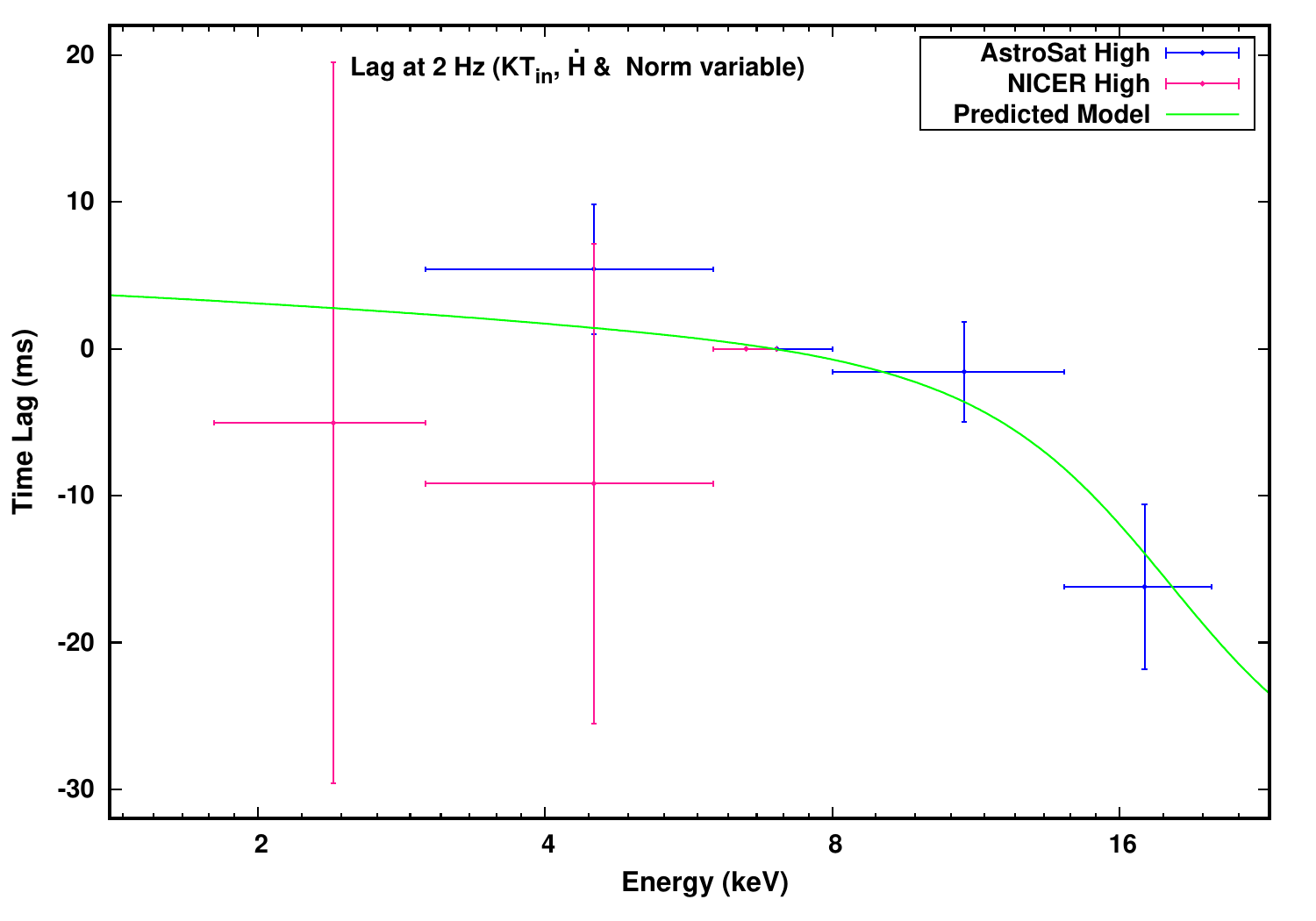}
        \label{fig:subfig5_2}
    }
    \par\medskip 
    \subfloat{
        \includegraphics[width=0.49\textwidth]{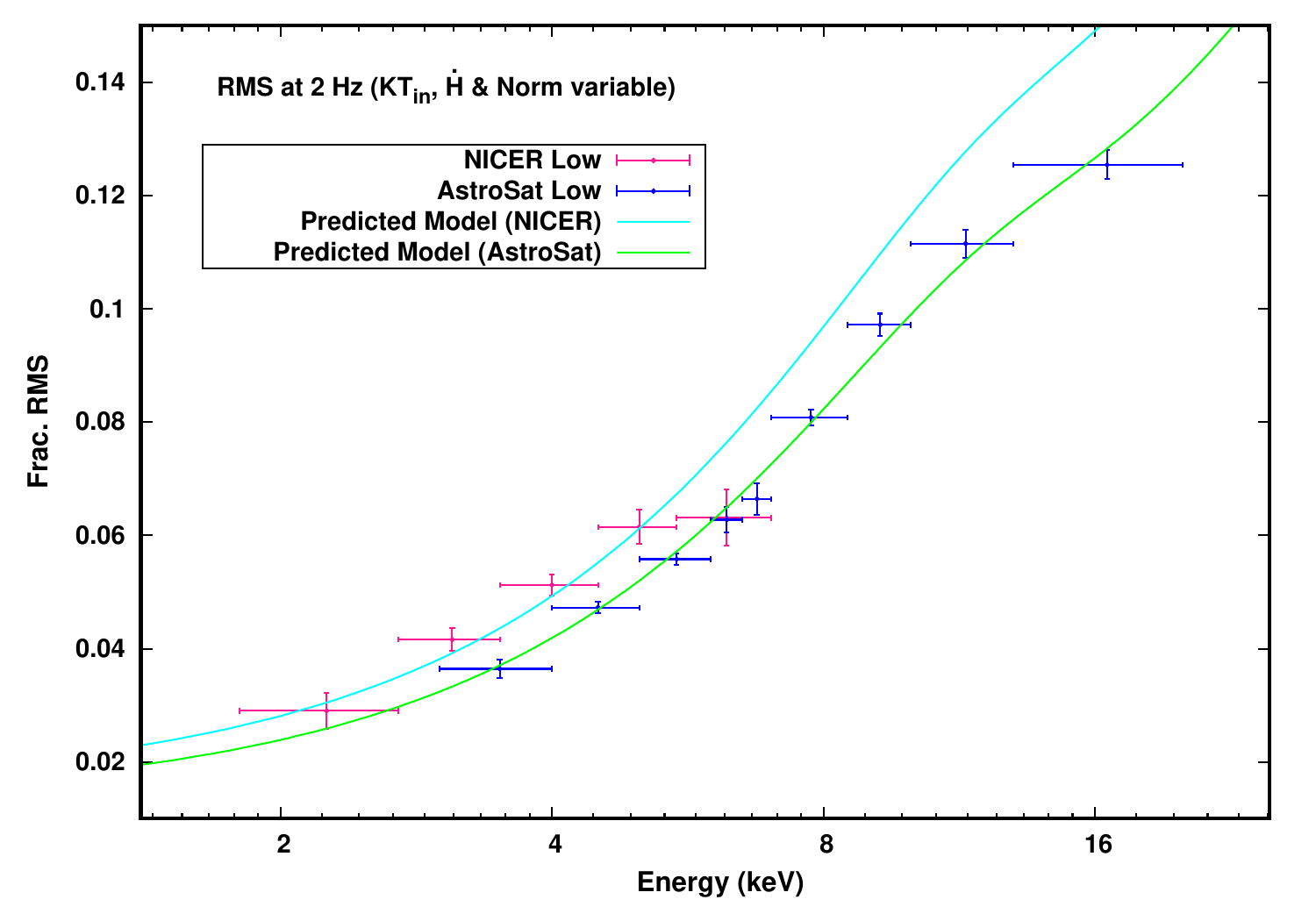}
        \label{fig:subfig5_3}
    }
    \hfill
    \subfloat{
        \includegraphics[width=0.49\textwidth]{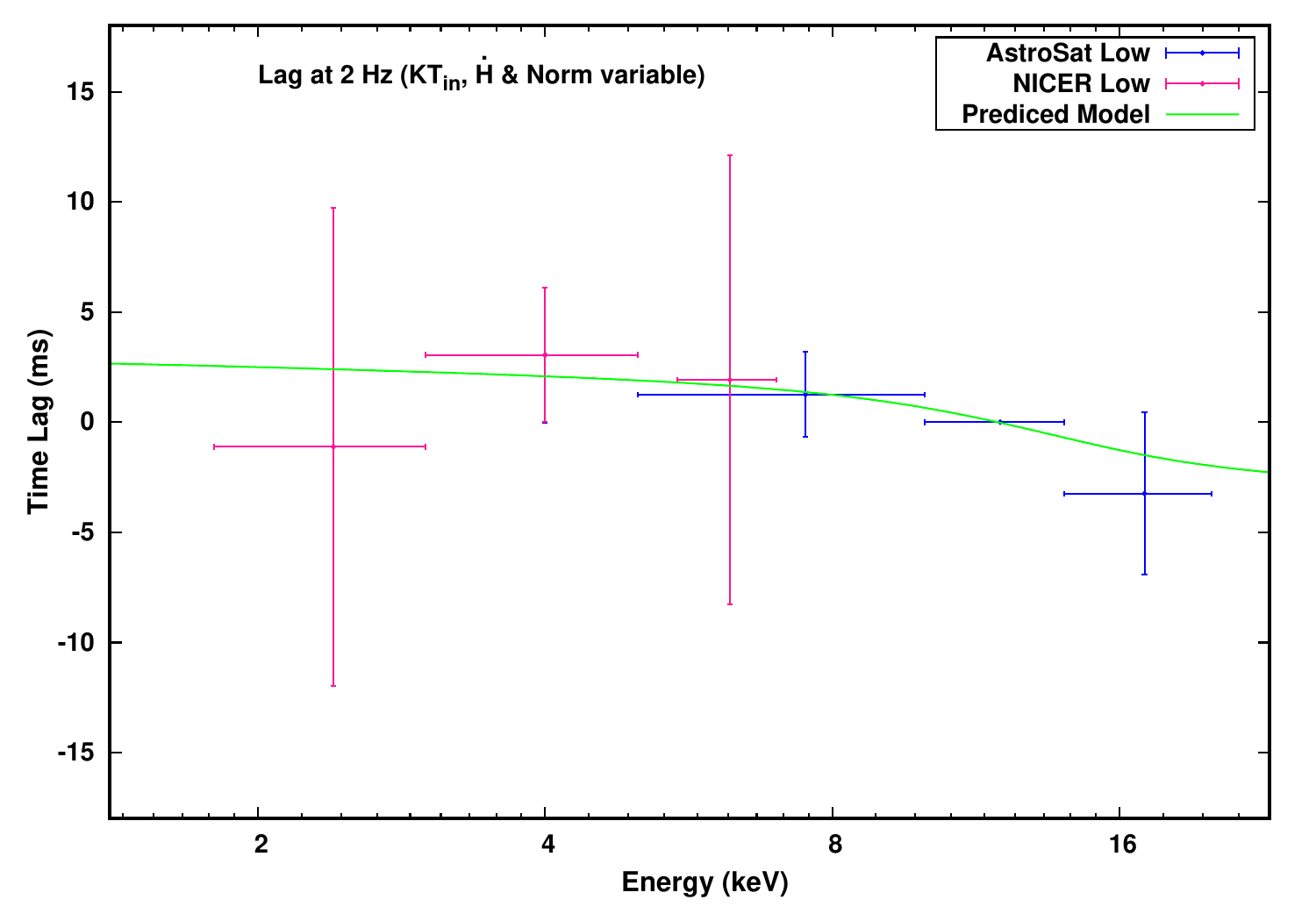}
        \label{fig:subfig5_4}
    }
   \caption{The top-left panel depicts the fractional RMS spectrum of AstroSat (For high flux at $\sim$2 Hz frequency), specifically LAXPC 10 and 20, along with NICER. Concurrently, the top-right panels present the time-lag spectrum for the high flux of AstroSat at approximately $\sim$2 Hz. The energy range used as a reference for calculating the time-lag spectra were chosen to be 6-8 keV band for LAXPC and 6-7 keV for NICER. On the other hand, the bottom-left panel shows the fractional RMS spectrum, while the bottom-right panel illustrates the time-lag spectrum for the low flux at around $\sim$2 Hz. In this case, the reference energy band used for calculating the time-lag spectra is 10.0-14.0 keV for LAXPC and 3-5 keV NICER. The solid line represents the predicted model for energy-dependent RMS and time lag, while the points show the observational results. The heating rate, Diskbb norm, and temperature were varied to explain the frac RMS and time lag variation. }
        \label{fig:fig5}
\end{figure*}
To calculate the fractional RMS across various energy bands, we first extracted the PDS for each respective energy band. To determine the RMS of the broad feature $\sim$2 Hz, we took the square root of the normalization value of the Lorentzian.
To generate time-lag spectra, the LAXPC `laxpc\_find\_freqlag' subroutine is used, which incorporates input parameters like frequency resolution ($\Delta$f) and the specific frequency (f) at which the time-lag has to be computed. The software 'laxpc\_find\_freqlag' has been described in \citet{yadav2016astrosat} and uses the standard technique to compute time-lag and its error as given in \citet{vaughan1997x, nowak1999rossi}. Similarly, for NICER fractional RMS values, we employed the `nicer\_find\_freqlag' subroutine. 
The phase lag is the time lag multiplied by $2\pi$ the frequency. Subsequently, we study the behavior of time-lag and fractional RMS in the frequency range 1.37-1.91 Hz for low and 1.91-2.45 Hz for high flux regimes. In Figure \ref{fig:fig5}, the top-left panel shows the fractional RMS spectrum of LAXPC (for high flux at $\sim$2 Hz frequency), particularly LAXPC 10 and 20, along with NICER data. The shift in the RMS of LAXPC and NICER may be due to the dead time of LAXPC. Concurrently, the top-right panel in Figure \ref{fig:fig5} presents the time-lag spectrum for the high flux of AstroSat and NICER at $\sim$2 Hz frequency in PDS.  The reference bands for time-lag calculations were chosen to be 6-8 keV for LAXPC and 6-7 keV for NICER.  The bottom-left panel illustrates the fractional RMS spectrum, while the bottom-right panel portrays the time-lag spectrum for the low flux state at $\sim$2 Hz frequency in PDS. The reference energy band for computing time lags is 10.0-14.0 keV for LAXPC and 3-5 keV NICER. The NICER time-lags have been shifted such that the reference band (4–8 keV) matches with the time-lag found in 4–8 keV in LAXPC. The solid line corresponds to the predicted model depicting energy-dependent RMS and time lag, whereas the data points represent observed results. Two distinct solid lines represent the model for NICER and AstroSat fractional RMS. 
A constant has been included to account for the shift observed in the fRMS spectra of LAXPC and NICER, which may be attributed to the dead time of the LAXPC. Our observations indicate that fractional RMS increases as the energy increases for both low and high flux regimes. Additionally, we note that for the time lag linked with the broad feature at $\sim$2Hz for both the low and high flux, there is a clear pattern of the time lag decreasing as the energy level increases. \\
\begin{table*}
\centering
\caption{Best fit model parameters for describing observed fractional RMS and time lag variations. The parameter variations are denoted in percentage ($\%$). The last two columns represent the parameters for high and low flux segments.}
\label{tab:table4}
\begin{tabular}{c c c c c c} 
\hline \hline
Parameters & Description  & High flux & Low flux  \\  
 \hline  \\ [-7pt]
$\lvert \delta T_{in} \rvert$ & Variation in Disk temperature  & $3.0 \pm 0.4$  & $2.8 \pm 0.3$ \\ [4pt]
$\lvert \delta Norm_{dbb} \rvert$ &  Variation in Diskbb Norm & $-6 \pm 2$  & $-5\pm 1$ \\ [4pt]
$\lvert \delta \dot{H} \rvert$ &  Variation in Heating rate & $15 \pm 3$  & $17 \pm 1$ \\ [4pt]
$\phi_{1}$ & Phase lag of $\delta \dot{H}$ w.r.t $\delta T_{in}$ (radian) &
$-0.5 \pm 0.3$  & $-0.07 \pm 0.06$ \\ [4pt]

\hline
$\chi^2$/dof &  &  13.76/11 & 16.11/12 \\ 
\hline
\end{tabular}
\end{table*}
\subsection{Modelling the Energy-Dependent Timing properties}
Here, we model the fRMS and time-lag spectra of the broad feature we observed at $\sim$2Hz in PDS. First, we briefly discuss the technique developed by \citet{garg2020identifying} to model the fractional RMS and time-lag spectra using spectral information. They developed the model considering a simple picture of the accretion flow geometry where, in hard state, the thin and optically thick disk is truncated at a certain radius far from the ISCO, and its inner part is converted into a hotter, optically thinner medium of high-energy electrons known as hot inner flow, or corona. The physics behind this conversion of a disk into a hot medium is not yet clear but is sometimes associated with the evaporation of the disk in inner regions \citep[][]{meyer1994accretion}. \citet{garg2020identifying} propose that the QPO originates inherently from fluctuations in the physical spectral parameters of the components within the accretion flow. The heating rate is determined as the energy difference between the total emitted photons and the input seed photons (For a more comprehensive explanation, refer to \citet{garg2020identifying}).
According to their methodology, changes in spectral parameters trigger changes in the steady-state spectrum F(E). This relationship can be mathematically expressed as:
\begin{equation} 
    \Delta F(E) = \sum_{j} \frac{\partial{F}}{\partial{\alpha_{j}}}  \Delta \alpha_{j}
\end{equation}
wherein $F(E)$ signifies the time-averaged spectrum and $\Delta F(E)$ represents the change in spectra attributed to changes in spectral parameters $\alpha_{j}$. These variations of spectral parameters, $\Delta \alpha_j$ are treated as complex numbers to incorporate the phase lags, and they can be expressed as $\Delta \alpha_j = A_j \exp^{i \omega_j t}$, where $A_j$ and $\omega_j$ represent the amplitude and frequency of the sinusoidal variation, respectively. This approach defines the fractional RMS variation as $\frac{|\Delta F(E)|}{\sqrt{2} \times F(E)}$. The phase lag between two energies, $E_1$ and $E_2$, is given by the phase of the cross-correlation function $\Delta F^*(E_2)\Delta F(E_1)$. The phase lag between two signals is equal to the time lag multiplied by 2$\pi$ times the frequency of the signal. The observed lags and RMS variations can arise across any pair of spectral parameters. This model takes the coronal heating rate  ($\dot{H}$) as a spectral parameter in place of the coronal electron temperature. \\
In its present form, the prescription incorporates only the possibility of disk emission, which is comptonizied by a thermal plasma or specifically the xspec model combination {\fontfamily{pcr}\selectfont thcomp*diskbb}. In Section \ref{sec:spectral}, it has shown that the spectra for both high and low flux can be represented by a single Comptonization model, provided the analysis is restricted to photons with energy < 20 keV. Since the fRMS and time-lags estimated above (in section 5.3) are also for energies less than 20 keV, we predict their energy behavior using the spectral parameters obtained in the spectral fit. Note that the Gaussian line does not substantially contribute to the spectra, and its omission will not affect the results. Moreover, the spectral parameters obtained from the SXT/LAXPC and NICER/LAXPC combinations were similar. Hence, we use the best fit spectral parameters obtained from the NICER/LAXPC fits in Table \ref{tab:table2} and check which parameter variations can produce the observed energy-dependent fRMS and time lag.   
 
First, we explore the possibility of variations exclusively in coronal parameters. In this case, we include variation in fractional scattering ($\lvert \delta f_{sc} \rvert$), heating rate ($\lvert \delta \dot{H} \rvert$), and Thomson optical depth ($\lvert \delta \tau \rvert$). Specifically, we look at cases where both $\lvert \delta f_{sc} \rvert$ and $\lvert \delta \tau \rvert$ exhibit phase lags with respect to $\dot{H}$. In this case, we identified a poor fit with a ${\chi^2}/{dof}$ value of about 36.8 for the high flux regime and approximately 58.2 for the low flux regime and with large variations of parameters inconsistent with the assumptions of the model. Consequently, this case suggests that the corona alone is not sufficient to explain the fRMS and time-lag spectra of the broad feature at $\sim$2 Hz. Next, we considered the possibility of variation of the disk temperature ($\lvert \delta T_{in} \rvert$) and coronal heating rate ($\lvert \delta \dot{H} \rvert$) with a phase difference between the two. We observe an improvement in the fitting of both the fRMS and lag spectra, reflected in the ${\chi^2}/{dof}$ values to 4.8 for high flux and 4.7 for low flux. However, despite this improvement, the fit is still inadequate. Thus, we included the possibility that the normalization of the disk (or, equivalently, the inner disk radius) also varies. This case provides an acceptable fit, as shown in Table \ref{tab:table4}. Note that we have introduced a negative sign for the variation in disc normalization, which is equivalent to a $\pi$ phase shift or in other words, that there is an anti-correlation between the disk radius and temperature with no time lag. Figure \ref{fig:fig5} shows the fitted model, with the left panels illustrating the fractional RMS spectra for high (top left figure) and low flux (bottom left figure). The fitted time lag spectra for both high and low flux are shown in the right panels, with the top figure representing high flux and the bottom representing the low flux regime.

\section{CONCLUSION}
\label{sec:conc}

In this work, we conducted a spectral and timing study of available simultaneous AstroSat and NICER observations of GRS 1915+105 made on 31 Oct and 1 Nov 2017. The exposure time of the source for AstroSat is 15.3 ks, and for NICER, it is 4.89 ks on 31 Oct and  3 ks on 1 Nov 2017. Given the temporal variability of the source during the observation period, we segmented the lightcurve into two distinct parts, corresponding
to high and low flux levels. We studied the energy spectra and timing characteristics of the distinct flux levels. We fitted the spectra with an absorbed disk model with thermal comptonization and a broad iron emission line, represented by {\fontfamily{pcr}\selectfont tbabs*gabs*gabs(thcomp * diskbb + gauss)} and found that the model describes the spectra for energies upto 20 keV. We replaced the xspec model {\fontfamily{pcr}\selectfont diskbb} with the relativistic component {\fontfamily{pcr}\selectfont kerrd} and obtained a similar fit. Both combinations of LAXPC with SXT and LAXPC with NICER provided similar spectral parameters. The spectral fitting revealed that the primary difference between the low and high flux levels seems to be the disk component. In high flux, the temperature of the disk is higher, and its normalization is smaller compared to the low flux

The simultaneous observation by Astrosat and NICER allowed us to study the rapid temporal behavior of the source in  1.8-20 keV energy range for time-lag and RMS spectra.
The PDS generated from both Astrosat and NICER data for the low and high flux levels did not reveal the presence of the QPO but instead identified a broad feature. Our analysis focused on the broad feature at $\sim$2 Hz, where the power is notably prominent compared to other features in the PDS. The fractional root mean square (fRMS) and time-lag spectra for the feature at $\sim$2 Hz were generated for each flux level, showing that for both cases, the broad feature exhibited soft lags. 

To better understand the radiative mechanism behind the variability, we modeled the energy-dependent fRMS and time-lag of the feature at $\sim$2 Hz with the model put forward by \citet{garg2020identifying}. According to the model, the energy-dependent properties of the variability can be described with correlated variations of certain spectral components with respective time delays. 

Initially, we consider variations only in coronal parameters. This case includes variations in coronal parameters such as scattering fraction ($\lvert \delta f_{sc} \rvert$), heating rate ($\lvert \delta \dot{H} \rvert$), and Thomson optical depth ($\lvert \delta \tau \rvert$), with phase lags of $\lvert \delta_{sc} \rvert$ and $\lvert \delta \tau \rvert$ relative to $\dot{H}$. However, this case results in a poor match with observed fRMS and time lag, suggesting that the corona alone is insufficient to explain the fractional root mean square (fRMS) and time-lag spectra. 

Next, we considered the possibility that the inner disk temperature and the heating rate vary, but this did not provide an adequate fit to the data. Finally, considering variations in three spectral parameters, namely the inner disk temperature, the inner disk radius, and the coronal heating rate, resulted in a good fit to the energy-dependent fRMS and time lag for both the low and high flux levels. The inner disk radius and temperature variations were found to be anti-correlated to each other, with no phase lag between them. In contrast, the phase lag for the coronal heating variation is negative, implying that the coronal heating variation occurs first, and then the disk reacts.

The primary results of this work are that during the simultaneous observation by NICER and AstroSat of GRS 1915+105, the source showed two flux levels, high and low. The spectra for both levels could be fitted by a model consisting of disk emission and thermal comptonization of the disk photons for energies less than 20 keV. The spectral parameters obtained from considering NICER/LAXPC were consistent with the ones obtained from SXT/LAXPC. The fractional RMS increases with energy, which implies that the variability is dominated by the corona. However, fitting the energy-dependent RMS indicates that the variability is not only due to the corona but also significant variability in the disc. A similar result was obtained for C-type QPOs for GRS 1915+105 by \citet[]{garg2020identifying} and for MAXI 1535–571 by \citet[]{garg2022energy} using only AstroSat/LAXPC data. This work shows that it is true when considering low-energy NICER data and the broadband noise for GRS 1915+105.

The work highlights the need for longer simultaneous observations of the source with AstroSat and NICER. Such observations would have the statistics to get well-constrained time-lags, especially in the NICER energy bands, and hence would provide better estimates of model parameters. We have not included reflection components in the spectral fitting for simplicity, as the timing model does not account for reprocessing effects. Incorporating the reflection components in spectra can be vital for future studies, ensuring a more precise representation. Importantly, the timing model utilized in this study requires further enhancement to accommodate more sophisticated spectral models, ensuring a more detailed comparison with high-quality data. \\

\section*{Acknowledgements}
We thank the referee for the useful comments, which have improved the manuscript significantly. We would like to express our gratitude for the assistance provided by the Indian Space Research Organization (ISRO) in facilitating mission operations and disseminating data via the ISSDC. The present study used data from the Soft X-ray Telescope (SXT) developed at the Tata Institute of Fundamental Research (TIFR) in Mumbai. The authors acknowledge thankfulness to the SXT Payload Operations Centre (POC) at TIFR for their valuable assistance in verifying and releasing the data via the International Space Science Data Centre (ISSDC) data repository, as well as for providing the necessary software tools required for data analysis. The LAXPC POC, TIFR, Mumbai, is also acknowledged for providing the necessary data analysis tools. This research has used data, software, and web tools obtained from the High Energy Astrophysics Science Archive Research Center (HEASARC), a service of the Astrophysics Science Division at NASA/GSFC. The author, RD, acknowledges the frequent visits to the Inter-University Centre for Astronomy and Astrophysics (IUCAA) in Pune to carry out the major part of the work.

\section*{Data Availability}
The observational data used in this paper are publicly available at ISRO’s Science Data Archive for AstroSat Mission (\url{https://astrobrowse.issdc.gov.in/astro_archive/archive/Home.jsp}) and  NASA’s High Energy Astrophysics Science Archive Research Center (HEASARC; \\ (\url{https://heasarc. gsfc.nasa.gov/})) and references are mentioned.

\vspace{5mm}


\end{document}